\documentclass{article}
\usepackage[utf8]{inputenc}
\usepackage{graphicx}
\graphicspath{ {./images/} }
\usepackage{enumitem}
\usepackage{url}
\usepackage{nameref}

\usepackage[dvipsnames]{xcolor}
\usepackage{hyperref}

\colorlet{mylinkcolor}{violet}
\colorlet{mycitecolor}{YellowOrange}
\colorlet{myurlcolor}{Aquamarine}
\hypersetup{
    breaklinks=true,
	colorlinks=true, 
	linktoc=all,     
    linkcolor=violet,
    filecolor=magenta,      
    urlcolor=cyan,
    citecolor=YellowOrange
}

\usepackage{amsthm}
\theoremstyle{definition}

\usepackage{geometry}
\newgeometry{vmargin={35mm}, hmargin={35mm,35mm}}

\usepackage{listings}
\usepackage{placeins}
\usepackage{titlesec}
\usepackage{titletoc}
\usepackage[title,titletoc]{appendix}

\usepackage{caption}
\usepackage{array,tabularx}

\usepackage[skins]{tcolorbox}
\newtcolorbox{protocolframe}[2][]{%
  enhanced,colback=white,colframe=black,coltitle=black,
  sharp corners,boxrule=0.4pt,
  fonttitle=\itshape,
  attach boxed title to top left={yshift=-0.3\baselineskip-0.4pt,xshift=2mm},
  boxed title style={tile,size=minimal,left=0.5mm,right=0.5mm,
    colback=white,before upper=\strut},
  title=#2,#1
}
\newcommand{\subsubsubsection}[1]{\paragraph{#1}\mbox{}\\}
\usepackage{fancyvrb}

\usepackage{chngcntr}
\counterwithin{figure}{section}

\title{\bfseries{Latus Incentive Scheme: Enabling Decentralization in Blockchains based on Recursive SNARKs}}

\author{
    Alberto Garoffolo\\
    \texttt{alberto@horizen.global}\\
    \texttt{Horizen}
    \and
    Dmytro Kaidalov\\
    \texttt{dmytro.kaidalov@iohk.io}\\
    \texttt{IOHK Research}
    \and
    Roman Oliynykov\\
    \texttt{roman.oliynykov@iohk.io}\\
    \texttt{IOHK Research}\\
    \texttt{\& V.N.Karazin Kharkiv National University}
}

\date{March 2021}

\begin{document}

\maketitle
\thispagestyle{empty}

\section*{ \centering }
\begin{center}
    \textbf{Abstract}
\end{center}
 In \cite{GKO20} we introduced a novel SNARK-based construction, called Zendoo, that allows Bitcoin-like blockchains to create and communicate with sidechains of different types without knowing their internal structure. We also introduced a specific construction, called Latus, allowing creation of fully verifiable sidechains. But in \cite{GKO20} we omitted a detailed description of an incentive scheme for Latus that is an essential element of a real decentralized system. This paper fills the gap by introducing details of the incentive scheme for the Latus sidechain. Represented ideas can also be adopted by other SNARK-based blockchains to incentivize decentralized proofs creation.

\newpage
{
  \hypersetup{linkcolor=black}
  \tableofcontents
}
\newpage
\section{Introduction}

Since emergence of Bitcoin \cite{N08} in 2008, decentralized ledger technologies were widely discussed by experts in various fields. Bitcoin became the first decentralized payment system. Absence of a centralized control over the network is claimed to be a disruptive feature that would allow more robust, fair, and transparent financial systems. Bitcoin inspired development of many other systems based on the same principle of decentralization with a variety of different features.  

With increased use of Bitcoin and other cryptocurrencies, their limitations became apparent: limited throughput, increased latency, reduced ability to scale and expand functionality, etc.\cite{C16}. 
In 2014, A.Back et al. proposed a concept of sidechains \cite{BCDF14} that overcomes the problems of a monolithic blockchain system. The basic idea is to enable creation of many blockchains with different functionalities that would be interoperable with the main blockchain. Interoperability in this case means ability to transfer a mainchain native asset (e.g., bitcoins) to and from a sidechain. In this way, a blockchain system, such as Bitcoin, may be extended with additional functionality implemented in a sidechain (e.g., smart contracts 
\cite{RSK}) that would use the same native asset, hence remaining in the Bitcoin ecosystem.   

In our previous paper \cite{GKO20} we introduced \textit{Zendoo} - a universal construction for Bitcoin-like blockchain systems that allows the creation of sidechains of different types without knowing their internal structure (e.g. what consensus protocol is used, which types of transactions are supported etc.), and the communication with such sidechains. Moreover, we proposed a specific sidechain construction named \textit{Latus} that can be built on top of this infrastructure, and would realize a decentralized verifiable blockchain system. We leveraged utilization of recursive composition of SNARKs \cite{Coda} to generate succinct proofs of sidechain state progression that are used to facilitate cross-chain transfer of coins between the mainchain and a sidechain.  

Even though in \cite{GKO20} we rigorously described the cross-chain communication protocol and the consensus protocol for the Latus sidechain, we put aside an incentive scheme. In this paper, we fill this gap and introduce incentives for the Latus sidechain participants. Given that the sidechain is not required to possess its own native asset or to mint new coins, it becomes non-trivial to construct a scheme that will hold in balance the various parties involved in its operation. In our paper, we identify such parties and their roles, and present a reward distribution scheme based purely on transaction fees, incentivizing them to follow the prescribed protocol to maximize overall system efficiency.   

The paper is structured in the following way: section 2 provides a brief overview of the sidechain construction presented in \cite{GKO20} and gives a more detailed description of some components that are important for the incentive scheme. Section 3 provides a description of the incentive scheme itself.  

\subsection{Related Work} 

A reward scheme that incentivizes honest protocol execution and decentralization is one of the most important parts of a blockchain system. Despite a certain amount of research in this area (e.g., \cite{ AB18}, \cite{LLP20}, \cite{ERZJ19}), there are still many open questions on how to build a robust incentive scheme that would facilitate honest behavior of participants.  

The first incentive scheme in a decentralized blockchain system was introduced in Bitcoin \cite{N08}. It is based on rewards for miners who are producing new blocks. The reward consists of newly minted coins and transactions fees. Even though the scheme was criticized for its vulnerability to protocol deviation attacks (such as selfish mining \cite{ES14}) and its tendency to centralization via creation of mining pools \cite{AW18,GKCC14}, it is still in use and secures the Bitcoin network without any major disruptions. A similar approach was reused by many other blockchain systems. 
 
A step further towards solving problems present in the Bitcoin incentive mechanism was made in \cite{BKKS20}. The paper studies reward sharing schemes in proof-of-stake blockchain systems that involve a large number of stakeholders. It is focused on the fair formation of stake pools that promotes decentralization. 
 
A big difference between our solution and the mentioned schemes is that we consider only transaction fees as the source of rewards without relying on inflation of new coins. 
 
In \cite{CST20}, the authors discuss the economics and monetary policy of the Mina\footnote{Formerly known as Coda.} protocol. Given that the Latus sidechain construction was inspired by the Mina design (in terms of constructing succinct proofs of state transitions using zk-SNARKs), there are also some similarities in the roles performed by system participants and reward distribution. As in Mina, our protocol has special entities called provers (analog to ``snarkers'' in Mina) that generate SNARK proofs necessary for transaction processing. Similarly, provers establish a market for proofs competing with each other to provide the cheapest proofs. Unlike Mina, our scheme provides additional mechanisms to parallelize the work among different provers in order to improve efficiency and to prevent certain types of deviation attacks. This foundation also defines incentives for other roles involved in the system. 
\section{Latus Sidechain Overview}

In \cite{GKO20}, we proposed \textit{Zendoo}, a construction that allows creation of  different sidechains, and communication with these sidechains without knowing their internal structure. We considered a parent-child relationship between the mainchain and sidechains, where sidechain nodes directly observe the mainchain, while mainchain nodes only observe cryptographically authenticated certificates from sidechain maintainers. Among other things, such certificates authorize transfers coming from sidechains. Certificate authentication and validation are achieved by using SNARKs \cite{BCTV13} that enable succinct proofs of arbitrary computations. The main feature of the construction is that sidechains are allowed to define their own SNARKs, thus establishing their own rules for authentication and validation. The fact that all SNARK proofs comply with the same verification interface used by the mainchain, enables great universality, as a sidechain can use an arbitrary protocol for authentication of its certificates. E.g., the sidechain may adopt a centralized solution, where the SNARK just verifies that a certificate is signed by an authorized entity (like in \cite{BCDF14} ) or, for instance, a decentralized chain-of-trust model as in \cite{GKZ19}.  

Latus is a sidechain built on top of the Zendoo's infrastructure. It realizes a decentralized verifiable blockchain system, and leverages utilization of recursive composition of SNARKs to generate succinct proofs of the sidechain state progression that are used to generate certificate proofs for the mainchain. This allows the mainchain to efficiently verify all operations performed in the sidechain without knowing any details about those operations.  

In this section, we give a brief overview of the Latus construction, and provide details on the parts directly related to the incentive scheme. For more details on Zendoo and Latus refer to~\cite{GKO20}. 

\subsection{Latus Consensus Protocol}

The consensus protocol is based on a modified version of the Ouroboros proof-of-stake
consensus protocol \cite{KRDO17}. Here we provide a short overview of the protocol, see more details in \cite{GKO20}.

\begin{figure}[htbp]
	\centering
	\includegraphics[trim={1cm 10.55cm 3cm 4.5cm}, clip,width=1\columnwidth] {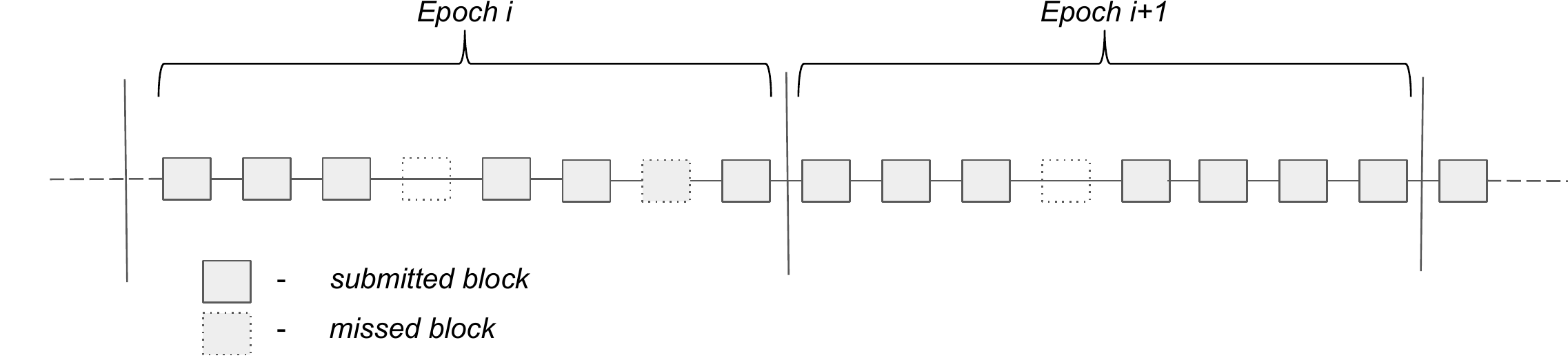}
	\caption{A general scheme of an epoch}
	\label{fig:2_1}
\end{figure}

In Ouroboros, time is divided into epochs with a predefined number of slots. Each slot is assigned with a slot leader who is authorized to generate a block during this slot. Slot leaders of a particular epoch are chosen randomly before the epoch begins, from a set of all sidechain stakers (Fig.~\ref{fig:2_1}). The protocol operates in a synchronous environment, where each slot takes a specific amount of time (e.g., 1 min).

\begin{figure}[htbp]
	\centering
	\includegraphics[trim={1cm 10.57cm 4cm 6.5cm}, clip,width=1\columnwidth] {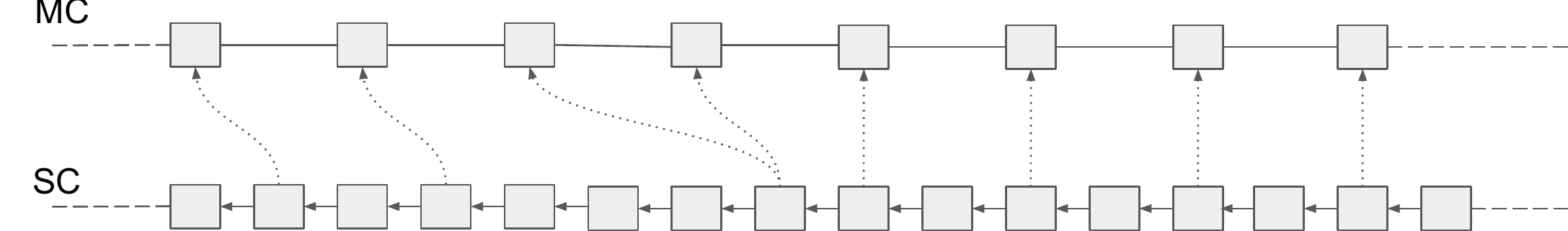}
	\caption {An example of the sidechain binding to the mainchain} 
	\label{fig:2_2}
\end{figure}

In our construction, we additionally introduce binding with the mainchain. Thus the sidechain blocks contain references to mainchain blocks, so that their history is preserved in the sidechain. The chain resolution algorithm is altered to enforce the sidechain following the longest mainchain branch
(Fig.~\ref{fig:2_2}).  

Even though it is not mandatory to include mainchain references, block forgers are incentivized to do this to earn rewards and collect fees from forward/backward
transfers.

Binding provides two important properties of our sidechain
construction:

\begin{enumerate} 
    \item \textbf{Deterministic synchronization between the mainchain and sidechain.} When the sidechain block ${SB}_i$ refers to the mainchain block $B_j$, it explicitly acknowledges all transactions included into the block $B_j$. It means that if $B_j$ contains any transaction related to this sidechain (e.g., forward transfers or backward transfer requests), such transactions are immediately included into the sidechain.
    \item \textbf{Mainchain forks resolution.} It is known that Nakamoto consensus does not provide finality on a chain of blocks \cite{SJS18}. It means that there is always a non-zero probability that some sub-chain of MC blocks will be reverted and substituted by another sub-chain with more cumulative work. Such behavior is normally handled by the mainchain, but may be disastrous for the sidechain, as $MC\rightarrow SC$ transactions that are already confirmed in the sidechain may be reverted in the mainchain. The binding eliminates such situations, because in the case of a fork in the MC, SC blocks that refer to forked blocks in the MC would also be reverted.
\end{enumerate}   

\subsection{Cross-Chain Communication}

The cross-chain transfer protocol allows communication between the mainchain and a
sidechain. It provides the ability to perform forward and backward transfers.

Forward transfer is an operation that moves coins from the original blockchain (the
mainchain) to the destination sidechain. On the mainchain side, it is implemented as a special type of transaction that destroys coins and provides metadata for creating coins in the sidechain. When the block with such a transaction is referenced by the sidechain, the transaction is synchronized and processed, issuing a corresponding amount of new coins to the receiver.
  
A backward transfer is an operation that moves coins from a sidechain to the mainchain. Normally\footnote{There is also a mainchain withdrawal request, see more info in \cite{GKO20}}, it is initiated on the sidechain as a special transaction that destroys coins. All backward transfer transactions submitted to the sidechain during a certain period -- called ``withdrawal epoch'' -- are collected in a special withdrawal certificate and pushed to the mainchain for processing. 
   
\begin{figure}[htbp] \centering
    \includegraphics[trim={1cm 9cm 4cm 5.42cm}, clip,width=1\columnwidth] {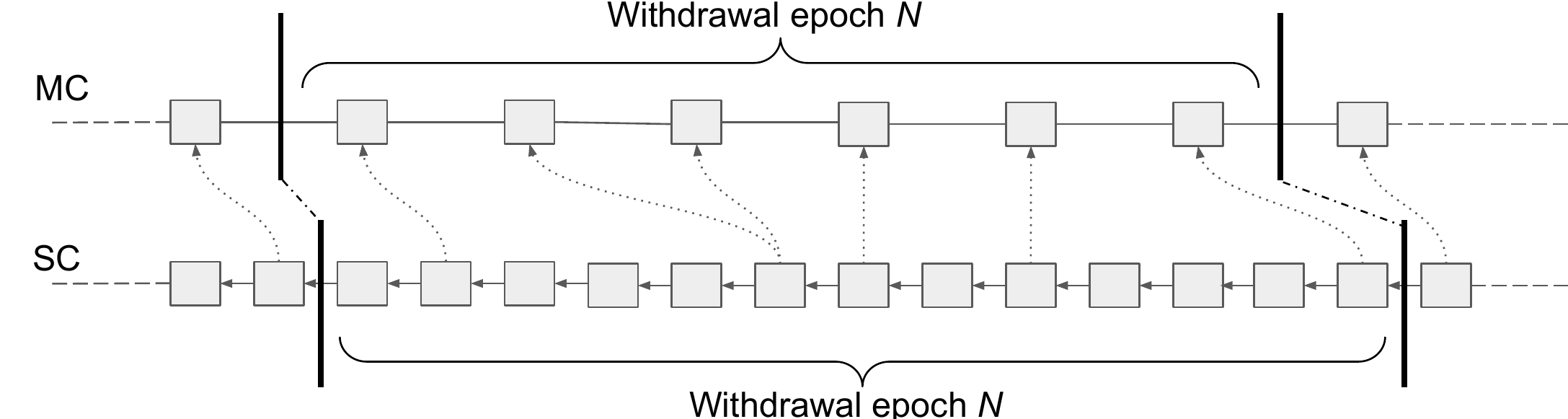}
    \caption{A withdrawal epoch in the sidechain is defined by a range of MC blocks}
    \label{fig:2_3}
\end{figure}

Withdrawal certificates are more than just a container for backward transfers, they are a kind of sidechain heartbeat that is submitted after the end of every epoch to the mainchain, even though there might be no backward transfers. A withdrawal epoch is defined by a range of MC blocks (see Fig. \ref{fig:2_3}).   
   
Importantly, a withdrawal certificate commits to a sidechain state and provides a SNARK proof of a correct sidechain state progression. This allows the mainchain to efficiently verify all operations performed in the sidechain without knowing any details about those operations. Particularly, it can safely process backward transfers included in the certificate and issue corresponding amounts of coins.   
   
\subsection{State Transition Proof}
   
\begin{figure}[htbp]
	\centering
	\includegraphics[trim={2cm 6.4cm 2cm 3cm}, clip,width=1\columnwidth] {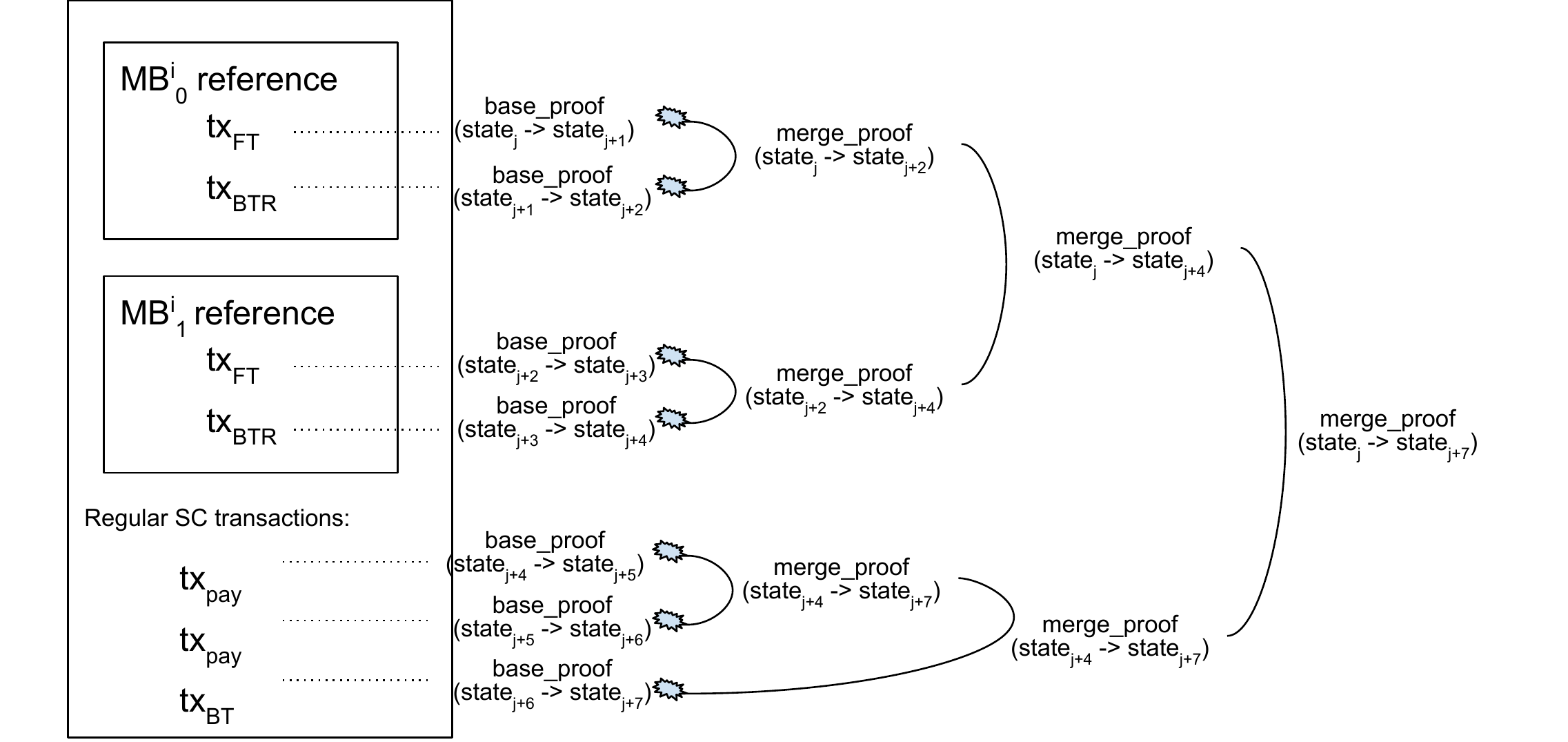}
	\caption{ Recursive composition of state transition proofs for the whole SC block. At the bottom level, there are proofs for basic transitions (represented by transactions included into the block) which are then recursively merged into a single proof} \label{fig:2_4}
\end{figure}

Each operation in the sidechain (including forward and backward transfer transactions) performs a state transition. We can consider all transactions in a block as a sequence of transitions that modifies the state obtained from the previous block. In \cite{GKO20} we defined how a sequence of state transitions can be proven using recursive SNARK compositions. The basic idea is that for each state transition a corresponding SNARK proof is generated, and then they are recursively merged into a single proof of state transition (see Fig.~\ref{fig:2_4}).
   
Each forger, when creating a block, includes a single proof of transition from the state before the block to the state after the block is applied. Eventually, there will be a single proof of state transition for the entire withdrawal epoch that will be included into the withdrawal certificate. For more details on state transition proofs, see section 5.4 in \cite{GKO20}.   
   
The remainder of this subsection discusses the distributed proof generation mechanism that is related to the incentive scheme.

\subsubsection{Distributed proof generation with proof substitution} \label{sec:distr_proof_gen}

Proof generation is a complex computational task. Provided with a set of $n$ transactions, a forger of the block will have to generate approximately $2n$ SNARK proofs to get a final block proof\footnote{Note that here and in other examples we assume a binary tree of proofs, where a base proof represents one transaction. The scheme can be optimized further, e.g., by proving several transactions with a single base proof, or by introducing some other merging strategy. In this case, the number of SNARK proofs per transaction may vary.}. Assuming that resources of the block forger are limited, that will inevitably reduce the number of transactions that can be processed. To improve throughput and for better utilization of idle resources of the network, we propose an additional mechanism for distributing proof generation. We refer to entities participating in the proof generation as \textit{provers}. Note that forgers also can (and in most cases will) be provers.  

\begin{figure}[htbp]
    \centering
    \includegraphics[trim={2cm 10cm 2cm 2cm}, clip,width=1\columnwidth] {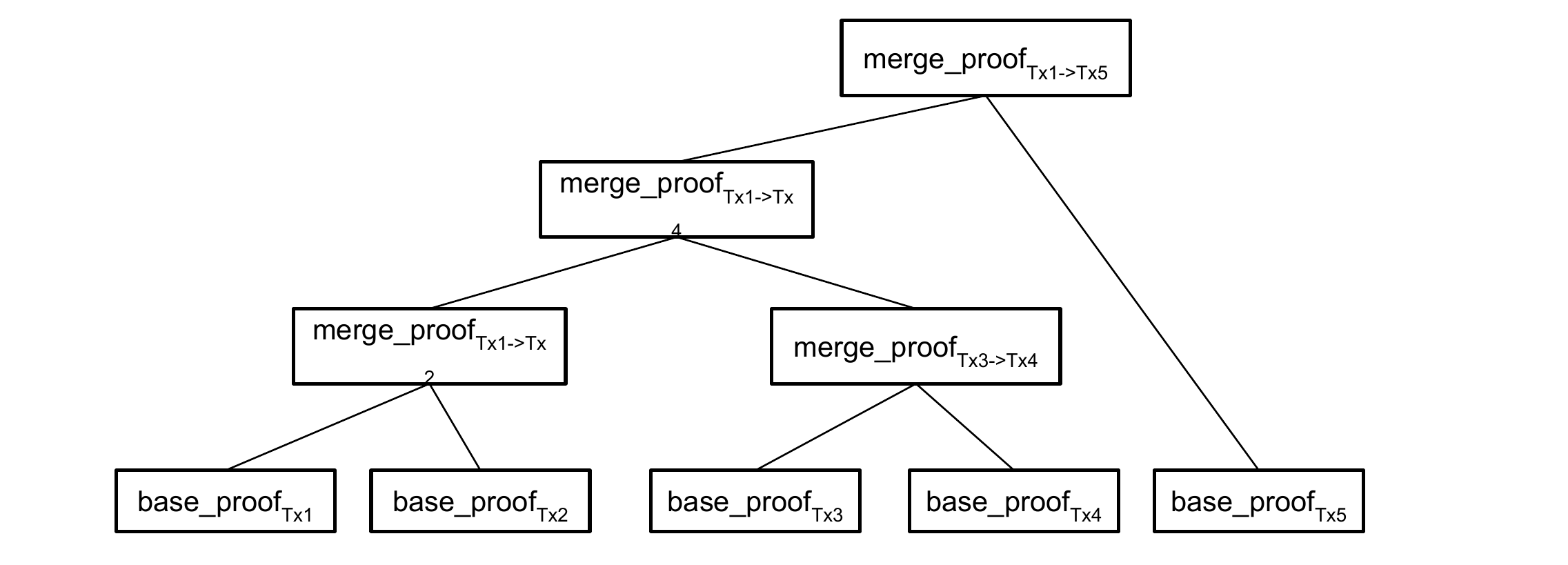}
    \caption{Tree of proofs for 5 transactions}
    \label{fig:2_5}
\end{figure}

The basic mechanism is as follows: a slot leader of the slot $s$, before issuing a block, broadcasts a message specifying identifiers of transactions he would like to include into the block (together with the upper bound of the reward that will be paid for a proof). We call this message a \textit{Transactions Proposal}. Given an ordered list of transactions, everyone can make up a tree of proofs that should be generated to obtain the block proof (see Appendix~\ref{sec:AppendixD} for details on tree construction). For example, if assuming a transactions proposal containing 5 transactions, the corresponding tree of proofs is shown on Fig.~\ref{fig:2_5}. 

The provers who are willing to participate in generation, can pick a certain proof and generate it for a self-set fee (within the upper bound set by the block forger). As provers decide themselves what proofs to generate, there might be duplicate proofs. To reduce duplication rate, there is an additional mechanism of random distribution based on ranking (discussed in the following section). All generated proofs are immediately broadcasted into the network, so that provers can continue working on upper level merge proofs that depend on lower level ones. While selecting proofs for merging, the rule is to take proofs with the lowest cost.

\begin{figure}[htbp]
    \centering
    \includegraphics[trim={2cm 9cm 2cm 3cm}, clip,width=1\columnwidth] {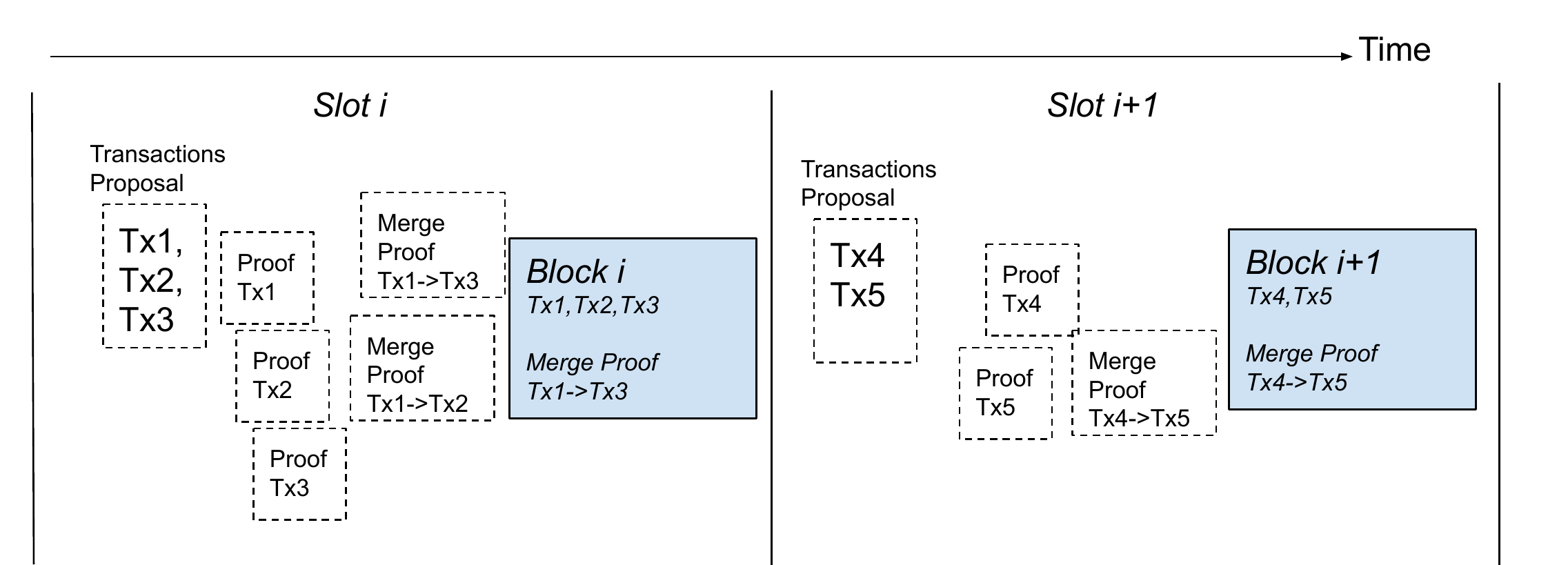}
    \caption{Distributed proof generation}
    \label{fig:2_6}
\end{figure}

Eventually, all generated proofs are collected by the block forger. He selects the tree of proofs that provides him with the highest total reward, and includes into the block the top level proof including information needed to reward every contributing prover. The scheme is visualized~on~Fig.~\ref{fig:2_6}. 

\FloatBarrier
\subsubsubsection{ Proofs substitution}  

Note that there can be several similar proofs for the same position in the tree, and it might happen that an already generated tree does not include the cheapest proofs (by the cheapest we mean the one with the lowest fee set by the prover). For instance, a cheaper proof might appear in the network with a delay, after provers already generated upper level proofs. In this case, a forger can pick up the cheapest proof, and include it into the block, “substituting” the one that was used to build a final merge proof. In this case, the reward will be paid to the prover of the cheapest proof, while the substituted one will receive nothing. Note that there is practically no regeneration of the final merge proof with a new intermediate proof, but
rather just an evidence is included into the block that a cheaper proof exists. Having such a mechanism of substitution is important for security reasons, as it reduces possible manipulations with the self-set fee for proofs. It also encourages provers to set reasonable fees for their proofs, as there will be a risk of being substituted. 

\begin{figure}[htbp]
    \centering
    \includegraphics[trim={2cm 12cm 12cm 2cm}, clip,width=0.6\columnwidth] {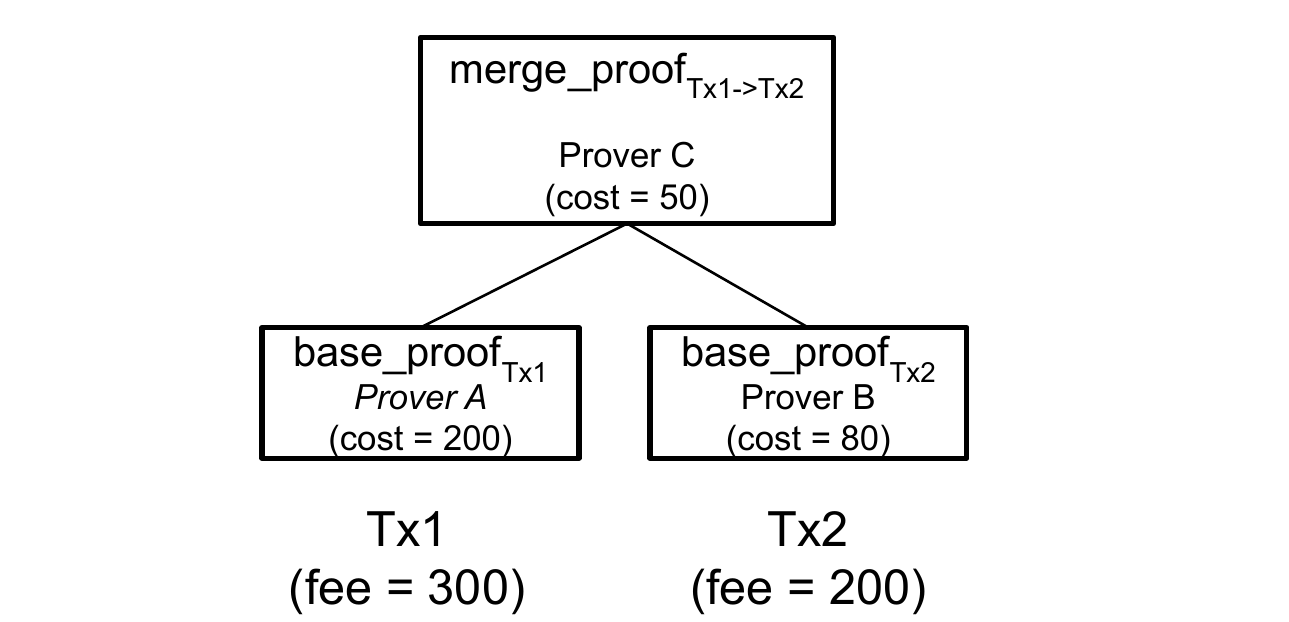}
    \caption{Tree of proofs}
    \label{fig:2_7}
\end{figure}

\textbf{Example}. Let us assume that a slot leader at the beginning of the slot issued a transaction proposal $\{Tx1,Tx2\}$. The provers quickly generated all the necessary proofs (see Fig.~\ref{fig:2_7}).    

Having $merge\_proof_{Tx1 \rightarrow  Tx2}$, the forger is already able to create a block and include transactions $Tx1$ and $Tx2$ into it, together with the proof. But let us assume that at the end of the slot, before the block is issued, there appears $base\_proof^{'}_{Tx1}$ from the prover \textit{D} with the lower cost $100$ that is half the price of the original $base\_proof_{Tx1}$ from the prover \textit{A}. The forger can include $base\_proof^{'}_{Tx1}$ along with the merge proof $merge\_proof_{Tx1 \rightarrow Tx2}$, in which case the prover \textit{D } will receive 100 coins, and the prover \textit{A } will receive nothing. Note that the merge proof is not regenerated and still includes the original proof $base\_proof_{Tx1}$. This makes substitution cheap, as it does not require regeneration of all upper level proofs.

\subsubsubsection{ Proofs usage and transition}

\begin{figure}[htbp]
    \centering
    \includegraphics[trim={3cm 9.5cm 2.5cm 1.5cm}, clip,width=0.98\columnwidth] {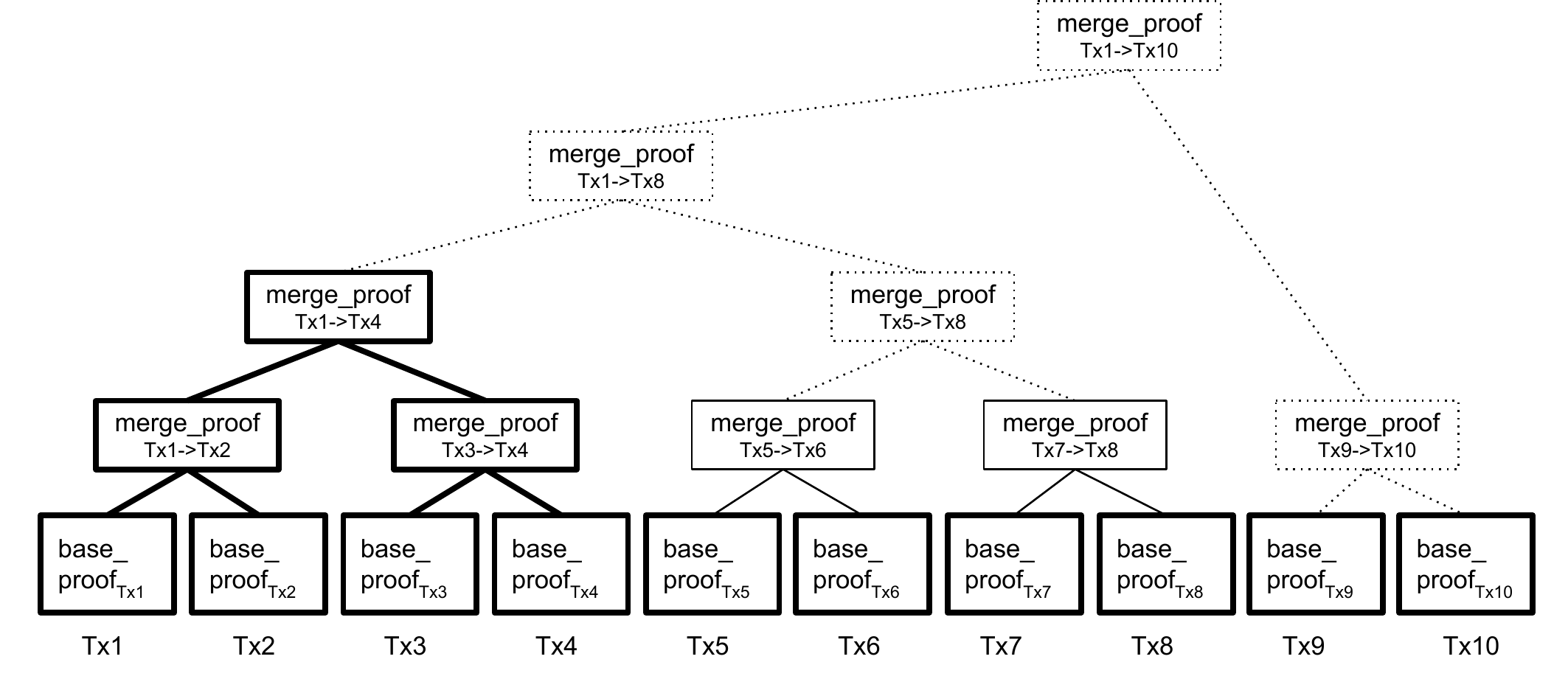}
    \caption{Proofs generation. Solid rectangles represent proofs that are already generated and broadcasted. Bold solid rectangles represent proofs that are ready to be used. Dotted rectangles signify proofs that are not yet generated.}
    \label{fig:2_8}
\end{figure}

If not all proofs are ready, the moment a forger issues a block, it should generate missing proofs or skip transactions that are not yet proven. If there are some already generated proofs that are not included into the current block (e.g., due to absence of upper level merge proofs), they can be reused by the next block forger.   Let us consider an example on Fig.~\ref{fig:2_8}. Assume that the forger at the beginning of the slot broadcasted a transactions proposal with ten transactions $\{Tx1,\ ...,\ Tx10\}$.

During the slot time, provers were not able to generate all the proofs required to include all ten transactions, but were able to construct a subtree of proofs only for transactions $\{Tx1,\ Tx2,$ $Tx3,\ Tx4\}$ (see Fig.~\ref{fig:2_8}).

In this case, the forger can include into his block only these transactions with $merge\_proof_{Tx1 \rightarrow Tx4}$. The next forger can construct a transactions proposal $\{Tx5,\ Tx6,\ ...,\ Tx10,\ ...\}$, in which case the already generated, but not used, proofs from the previous slot can be reused.  
 
Note that the forger can prepare all the proofs himself, in which case he may skip submitting a proposal. This mechanism is not enforced by the consensus protocol; rather, it is a layer-2 solution for improving throughput.   

\subsubsection{Rewards for provers}   
   
As previously mentioned, provers themselves set fees for generated proofs. The public key of a prover and requested fee are embedded into the proof itself, so that it is impossible for a forger to change these values. It is done in the following way: each proof (either base or merge) contains an additional public input of the form:  \[HP_{xy}=hash(HP_x\ |\ HP_y\ |\ xy\ |\ pk_{xy}\ |\ fee_{xy}),\]   
where  

\begin{itemize}
    \item[] $HP_x$, $HP_y$ are corresponding hash values of lower level proofs in the case this is a merge proof or $NULL$ in the case this is a base proof;
    \item[] $xy$ is the identifier of a particular transition $x \rightarrow y$ that is being proved;  
    \item[] $pk_{xy}$ is the public key of the prover; 
    \item[] $fee_{xy}$ is the  requested amount of fee. 
\end{itemize}

\begin{figure}[htbp]
    \centering
    \includegraphics[trim={2cm 9.5cm 3cm 4cm}, clip,width=1\columnwidth] {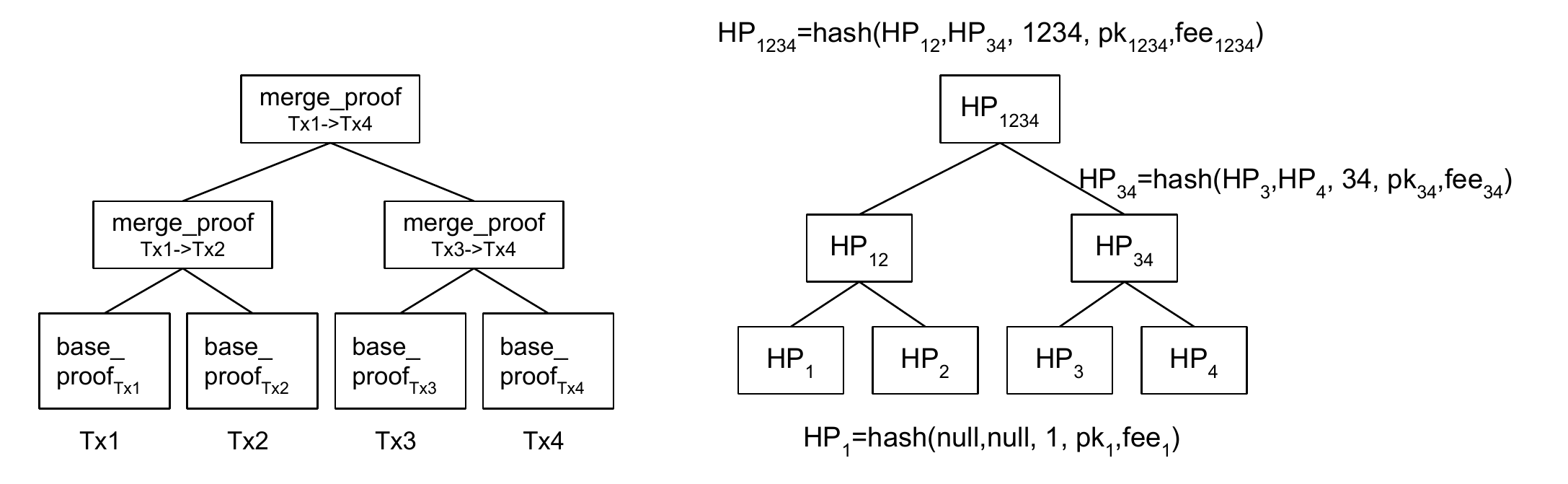}
    \caption{On the left, there is a Merkle tree of proofs, on the right there is the corresponding Merkle tree of provers. For every tree of proofs, there is a corresponding tree of provers that has the same form. The $H_x$ values from the provers tree serve as input when verifying corresponding proofs.}
    \label{fig:2_9}
\end{figure}

The block contains a set of all provers that contributed to the block proof with their public keys and amounts of requested fees. Verification of the block proof requires as a public input the value $HP$, which is the head of a Merkle tree with all intermediate $HP_i$. By reconstructing the $HP$ and checking it with the block proof, everyone can verify the provided set of provers. See example on Fig.~\ref{fig:2_9}. 

If the block contains substituted proofs, additionally it includes a set of substituting provers. For instance, let us assume that in the example on Fig.~\ref{fig:2_9}, two proofs were substituted with the lower cost proofs. In this case, the block will contain the following data:
\\~\\
\indent\textit{Block }\{ 
 
\setlength\parindent{34pt} Transactions: $Tx1$, $Tx2$, $Tx3$, $Tx4$ ;
 
State transition proof: $merge\_proof_{Tx1  \rightarrow Tx4}$;
 
Original provers: \{ 
  \setlength\parindent{44pt}
  
    $Prover_1$: $pk_1$, $fee_1$;   
  
    $Prover_2$: $pk_2$, $fee_2$; 
 
    $Prover_3$: $pk_3$, $fee_3$;   
    
    $Prover_4$: $pk_4$, $fee_4$;  
 
    $Prover_{12}$: $pk_{12}$, $fee_{12}$;   
 
    $Prover_{34}$: $pk_{34}$, $fee_{34}$;   
 
    $Prover_{1234}$: $pk_{1234}$, $fee_{1234}$; 
  \setlength\parindent{34pt} \}
 
Substituting provers: \{  
  \setlength\parindent{44pt}
  
    $Prover^"_{12}$: $merge\_proof^"_{Tx1 \rightarrow Tx2}$, $pk^"_{12}$, $fee^"_{12}$;   
 
    $Prover^"_4$: $base\_proof^"_{Tx4}$, $pk^"_4$, $fee^"_4$; 
  \setlength\parindent{34pt} \}  
  \setlength\parindent{24pt}
 
\}   
 
\subsubsection{ Provers selection and competition} 
    
 Given a transaction proposal and a corresponding tree of proofs that should be generated, we need to define a mechanism to distribute different proofs among different provers. The basic idea is that for each proof a prover will calculate his rank using a verifiable random function~\cite{MRV99}:
 
 \[{(rank}_i,\ p_{vrf})=VRF\_create(privKey,\ proofid_i),\]
 
 where $privKey$ is a private key of the prover and $proofid_i$ is a unique identifier of the SNARK proof to be generated. ${rank}_i$ is an output of the VRF function and $p_{vrf}$ is the proof of output correctness. Everyone can verify the rank using the public key of the prover: \[VRF\_verify(pubKey,\ {(rank}i,\ p{vrf})) \in \{ true, false \}.\]
 
     
 The VRF function is used to prevent provers from knowing ranks of other provers beforehand, so that they cannot adjust their strategy for proof selection. 
 
 The prover sorts proofs by ${rank}_i$ and starts generating them using ranking as priority index.  
 
 The block forgers and other provers eventually select proofs with the lowest cost or, if amounts are the same, they choose proofs with higher ranking. 
 
 Note that the goal of this mechanism is to distribute proofs randomly among provers. The ranking is not enforced by the consensus protocol. The provers can actually adopt other mechanisms for selecting proofs, but we assume that rational provers will adopt this strategy.    
 
\subsection{Withdrawal Certificate} \label{sec:wcert}
 
Withdrawal certificate is a standardized posting that allows sidechains to communicate with the mainchain. Its main functions are:

\begin{enumerate} 
    \item  Delivering backward transfers to the MC.
    \item  Serving as a heartbeat message enabling the MC to identify SC status.
\end{enumerate}  

A withdrawal certificate, to be included into the mainchain, should be appended with signatures from at least half of the forgers who issued blocks in the withdrawal epoch. The signatures are issued at the end of the epoch and collected by the certificate submitter to be sent together with the certificate to the mainchain. See more details on certificate signing in Appendix \ref{sec:AppendixA}. 

The certificate authorization and validation on the mainchain side relies on the included SNARK proof. The mainchain knows only the verification key -- that is registered upon sidechain creation -- and the interface of the verifier, that is unified for all sidechains. If the SNARK proof and public parameters are valid, then the certificate gets included and processed in the mainchain.

An important question is who should submit the certificate to the mainchain. Given that submission requires some work and fees to be paid on the mainchain, we should carefully consider the mechanism that is used to facilitate this process. The following section describes an incentivization mechanism for submitting certificates.

\subsubsection{Submission mechanism} \label{sec:submission_mechanism}

\begin{figure}[htbp]
    \centering
    \includegraphics[trim={0cm 9.5cm 2cm 2.7cm}, clip,width=1\columnwidth] {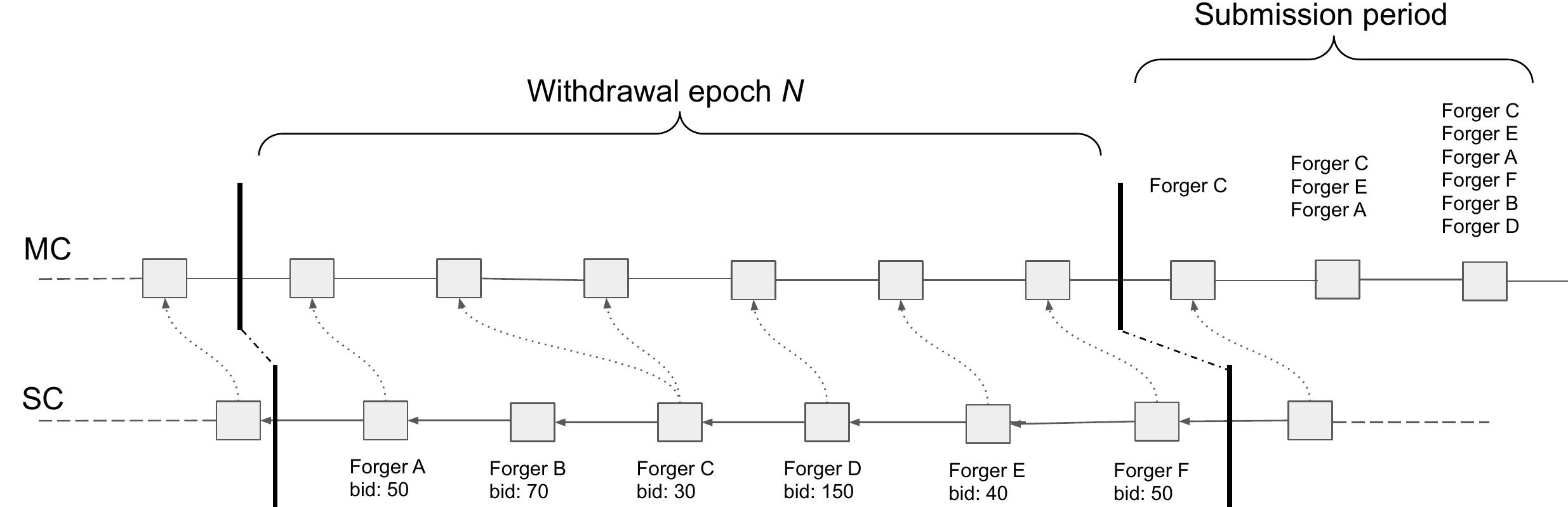}
    \caption{Withdrawal certificate submission priorities. With each new MC block, more forgers are allowed to submit a certificate}
    \label{fig:2_10}
\end{figure}

A withdrawal certificate can be submitted by anyone, but only forgers of the epoch may be rewarded for doing this. The forgers compete with each other for the right to be rewarded for submitting a certificate. It is done in the following way: while issuing a block, a forger includes a bid for the amount of coins he wants to receive as a reward for submitting the certificate. At the end of the epoch, all forgers are sorted in the ascending order by the amount of requested fees. The first forger entitled to submit is the forger with the lowest bid. To receive the reward, he should submit the certificate during a specified period. If he missed his opportunity, the next $N$ forgers ($N$ is a system parameter) would receive the right to be rewarded for submission and so on. On each step the number of forgers that can be rewarded is multiplied by $N$ to increase the chances that someone will eventually do this. The process is visualized on Fig.~\ref{fig:2_10}.

The forger will be rewarded with the requested fee only if he submits a certificate at a proper time according to priorities. Actually, MC will accept any valid certificate, but the submitter will not receive anything if he was not entitled to do so. See more information on submitter incentives in the following section.
\section{Incentive Scheme}

The ultimate goal of the incentive scheme is to establish a balanced mechanism of
reward distribution that facilitates stable system operation and honest participation. In this section we describe in detail the actors of the system that should be rewarded and operations for which they are rewarded, and provide explicit reward distribution formulas.

\subsection{Main Components}

Main components of the incentive scheme are defined by the types of operations available in the system. We identify the following operations that require some form of incentivization:  

\begin{enumerate} 
    \item  \textbf{Block forging} - participation in the sidechain consensus protocol, generating and issuing blocks.
    \item  \textbf{Mainchain block referencing} - including MC block references into SC blocks facilitating cross-chain communication.
    \item  \textbf{Transactions processing} - including SC transactions into blocks by forgers.
    \item  \textbf{Proofs generation} -  every transaction should be proven by a SNARK proof.
    \item  \textbf{Withdrawal certificate signing} - every forger should issue a signature for a withdrawal certificate at the end of the epoch.
    \item  \textbf{Withdrawal certificate submission} - at the end of a withdrawal epoch, a certificate with the current state of the sidechain should be generated and submitted to the mainchain.
\end{enumerate}

\noindent We define the following principal actors of the system that perform these operations:  

\begin{enumerate}
    \item  \textbf{Block forgers} - they do block forging, MC referencing, transactions processing, WCert signing and submission.
    \item  \textbf{Provers} - they do proof generation.
\end{enumerate}   

Note that in reality most likely (but not necessarily) these roles will be performed by the same physical entities (e.g., block forgers can also be provers).     
\\~\\
Moreover, we have two additional actors that we want to incentivize:
\begin{enumerate}
    \item  \textbf{Sidechain developer} - an entity that launched a particular sidechain instance.
    \item  \textbf{Circuit developer} - an entity that created a SNARK circuit whose verification key can be used  to validate withdrawal certificates of the sidechain. Note that different sidechains can use the same verification key.
\end{enumerate}   
       
These two entities are considered as sidechain "founders". Given that usually founders put a lot of effort into the initial development and maintenance of the system, we want to establish a mechanism that would allow them to be rewarded. The above incentivization is not strictly required for having a working system, and these rewards can be eliminated in a particular sidechain instance. The enforcement of such redistribution is not a part of the Zendoo protocol, but is managed by specific rules introduced by the circuit (e.g. the circuit can force redistribution of a percentage of the fees both to the sidechain developer and to the circuit developer, by enforcing inclusion of two specific backward transfer outputs into each withdrawal certificate, while a different circuit used by another sidechain might not offer any reward to either of the two actors).

\subsection{Source of Rewards}

Since there is no minting of new coins in the Latus sidechain, the only source of rewards are collected transactions fees. Note also that the reward distribution formula described in the next section, may be generalized and applied also to blockchains with other rewards sources.
     	 
\subsection{Basic Reward Distribution Formula}
     	   
Let us denote $Br_i$ as a total amount of fees collected in SC block $B_i$. Then, let $Er_j$ be the total amount of fees collected during the whole withdrawal epoch in the sidechain:  
\[ Er_j=Br_k+Br_{k+1}+...+Br_{k+n-1}, \]
that is the sum of the fees collected in all blocks of that epoch ($n$ is an epoch length).  
     	   
There are two categories of rewards: 
\begin{enumerate}
    \item  \textbf{Local rewards} - paid within a particular block from fees collected in block.  
   	\item  \textbf{Global rewards} - paid at the end of the epoch from a general pool of collected rewards.
\end{enumerate}  
   	
The fees of each block are split between these two categories:  \[Br_i={Br}^{lc}_i+{Br}^{gl}_i,\]
\[{Br}^{lc}_i=(1-gl)\cdot Br_i,\]
\[{Br}^{gl}_i=gl\cdot Br_i, \]
where $gl \in [0,1]$ is a system parameter defining percentage of block fees that goes into the global pool.    
   	
The \textbf{local rewards} are distributed in the following way:
\[{Br}^{lc}_i={Br}^{lc\_forger}_i+{Br}^{lc\_provers}_i,\]
where ${Br}^{lc\_forger}_i$ is the portion paid to the block forger and ${Br}^{lc\_provers}_i$ is paid to all provers who contributed to the state transition proof of this block. See more details about local rewards distribution in [\ref{sec:local_rewards_distr} \nameref{sec:local_rewards_distr}]. Note that if there are no transactions in the block, there would be no fees, thus, the block forger would not receive anything directly and nothing will be contributed to the global pool.
   	
The global rewards are collected into the pool and distributed at the end of the epoch. Let us denote ${Er}^{gl}_j$ as a global reward pool of the epoch \textit{j}. It comprises all ${Br}^{gl}_i$ belonging to the epoch:
\[{Er}^{gl}_j={Br}^{gl}_k+{Br}^{gl}_{k+1}+...+{Br}^{gl}_{n+k-1},\]
where $n$ is epoch length.
   	
The \textbf{global rewards} are distributed in the following way:
\[{Er}^{gl}_j={Er}^{gl\_forgers}_j+{Er}^{gl\_refs}_j+{Er}^{gl\_submitter}_j+{Er}^{gl\_scdev}_j+{Er}^{gl\_circuitdev}_j,\]
where each category takes a certain percentage of the global pool. Categories are as follows:
   	
\begin{itemize} 
    \item  ${Er}^{gl\_forgers}_j$ is divided among all forgers in the epoch proportionally to the number of blocks issued;
    \item  ${Er}^{gl\_refs}_j$ is divided among forgers that included mainchain block references (proportionally to the number of references);
    \item  ${Er}^{gl\_submitter}_j$ is paid to an entity that submitted a withdrawal certificate to the MC;
    \item  ${Er}^{gl\_scdev}_j$ is paid to a sidechain developer;
    \item  ${Er}^{gl\_circuitdev}_j$ is paid to a circuit developer.  
\end{itemize}  
   	  
More details about global rewards distribution are adduced in [\ref{sec:global_rewards_distr} \nameref{sec:global_rewards_distr}]. 
   	  
\subsection{Local Rewards Distribution} \label{sec:local_rewards_distr}
   	    
Local rewards are the portion of block rewards that are distributed among entities that are directly related to the production of a given block. There are two such entities:
   	    
\begin{enumerate} 
    \item  \textbf{Block forger} who issued the block.
    \item  \textbf{Provers} who contributed to generation of the state transition proof for this block.
\end{enumerate}
   	      
Let us assume a block has $n$ transactions, that, after deduction of global pool fees, provide ${Br}^{lc}_i$ fees to be distributed among the aforementioned entities. According to the proof generation mechanism, $n$ transactions require approximately $2n$ proofs\footnote{Again, in this example we assume a binary tree of proofs where a base proof represents one transaction. There can be other merging strategies requiring less proofs per transaction.} to be generated, so the provers portion of ${Br}^{lc}_i$ should be distributed among $2n$ entities:
\[{Br}^{lc}_i={Br}^{lc\_forger}_i+{Br}^{lc\_provers}_i,\]
\[{Br}^{lc\_provers}_i=\sum^{2n-1}_{p=0}{}{Br}^{lc\_provers(p)}_i,\]
where ${Br}^{lc\_provers(p)}_i$ is the reward paid for generating of a particular proof \textit{p}. 
   	       
The values ${Br}^{lc\_provers(p)}_i$ are set by the provers themselves while generating the proof \footnote{See [\ref{sec:distr_proof_gen} \nameref{sec:distr_proof_gen}] for more details}. The idea is that there will be an open competition to provide proofs with the lowest cost that will establish a fair market price for proofs.   
   	       
Given that ${Br}^{lc\_provers(p)}_i$ values are dynamic depending on the provers, it means that the reward for the block forger ${Br}^{lc\_forger}_i$ is also dynamic and made up of the residual of the local rewards after deduction of the provers rewards:
\[{Br}^{lc\_forger}_i={Br}^{lc}_i-\sum^{2n-1}_{p=0}{}{Br}^{lc\_provers(p)}_i.\]  
     
That provides an additional incentive for a forger to pick up proofs with the lowest cost.     
     
\begin{figure}[htbp]
	\centering
	\includegraphics[trim={0.8cm 6.4cm 2cm 1.3cm}, clip,width=1\columnwidth] {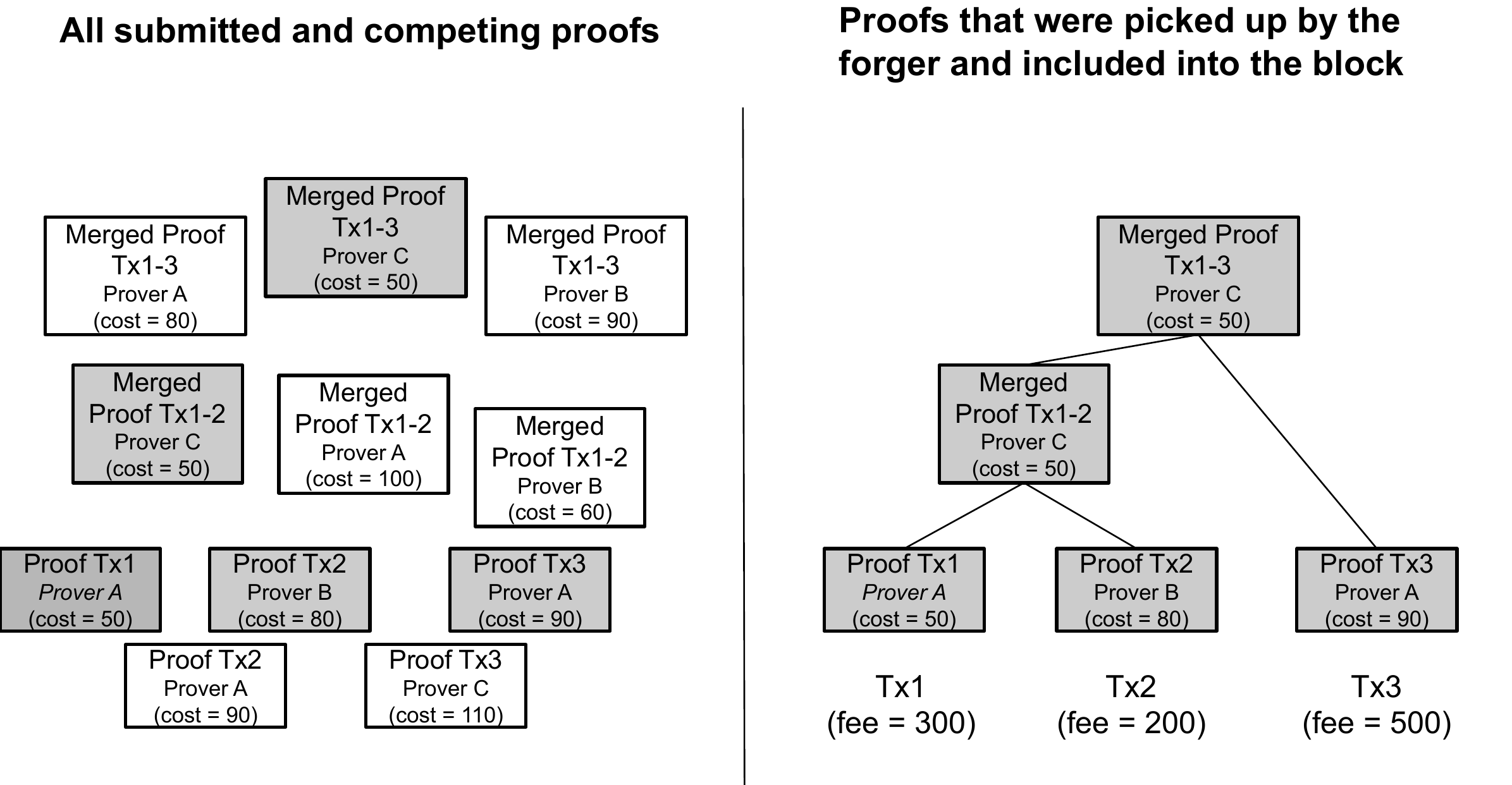}
	\caption{Example of proofs competition. Different provers generate proofs with certain costs. The cheapest proofs are picked up and used to produce the final merged proof.} \label{fig:3_1}
\end{figure}

Note that if there are no transactions in the block, there would be no local rewards (i.e., ${Br}^{lc}_i=0$). This is not a problem for state transition proof generation, because there is nothing to prove, so no work should be done by provers, but there is still work for the forger because he anyway has to issue a block. For this reason we introduce an additional reward for the forger, which is paid at the end of the epoch from the global pool (it is discussed in [\ref{sec:global_rewards_distr} \nameref{sec:global_rewards_distr}]). It incentivizes forgers to issue blocks even if they are empty.
\\~\\
\textbf{Example 1.} Let us assume we have 3 transactions in the txs pool which pays 300, 200 and 500 coins as a fee respectively. A forger that wants to issue a block with these transactions has to include a state transition proof that requires generation of 5 SNARK proofs that are eventually merged. Let us assume that we have three provers \textbf{A}, \textbf{B} and \textbf{C} competing to provide proofs. Each of them can submit a proof for each level. The forger will include the cheapest proofs as shown on Fig.~\ref{fig:3_1}.

As it can be seen from Fig.~\ref{fig:3_1}, the total amount of fees collected by the block is ${Br}_i=1000$ coins (the sum of all txs fees). Then, assuming the percentage that goes to the global pool is $gl=0.2$, the distribution of the rewards is as follows:
\[{Br}^{lc}_i=(1-gl) \cdot 1000=800,\]
\[{Br}^{lc\_provers}_i=50+80+50+90+50=320,\]
\[{Br}^{lc\_forger}_i={Br}^{lc}_i-{Br}^{lc\_provers}_i=800-320=480.\]    

\subsubsection{Local rewards for mainchain referencing}   

As we mentioned earlier, forgers are not obliged to include mainchain references into their blocks. The direct incentive to do this is the fees paid by cross-chain transactions included into referenced MC blocks (e.g. forward transfers). In this case, MC references are treated as the regular transactions that allow the forger to gain some fees. See more details about fees for cross-chain transfers in Appendix \ref{sec:AppendixB}.
   	         
As there can be empty mainchain block references when there are no cross-chain transactions, forgers who reference MC blocks earn an additional reward paid from the global pool.   

\subsection{Global Rewards Distribution} \label{sec:global_rewards_distr}
   	         
The global rewards are distributed after the end of a withdrawal epoch. The global reward pool is filled by a portion of the fees from each block in the epoch. It is used to incentivize entities to do some work that cannot be compensated by fees of a particular block. The recipients of global rewards are the following:
   	          
\begin{enumerate}
    \item  \textbf{Block forgers} - a portion of the global pool is distributed equally among all the forgers that issued blocks in this epoch. This is an extra reward for forgers in addition to what they have been paid directly from the local rewards. An additional incentive is important, as there may be blocks that do not have transactions at all and, thus, do not have local rewards. But it is very important for the consensus security to continue to issue blocks, even if empty. 

    Another part of the global pool is provided to the forgers which include mainchain block references into blocks.
    
    \item  \textbf{Withdrawal certificate submitter} - at the end of a withdrawal epoch, a certificate that commits to the state of the sidechain should be submitted to the mainchain. It requires additional cost that should be compensated to an entity performing submission.
    
    \item  \textbf{Sidechain developer} - we consider a sidechain developer to be an entity that launched and maintains a particular sidechain.
    
    \item  \textbf{Circuit developer} - the rules of a sidechain are enforced by a particular SNARK circuit, which is registered in the mainchain upon sidechain creation. SNARK development is a complex job that should be compensated. Note that the same circuit can be reused for different sidechains separating a sidechain developer from a circuit developer.
\end{enumerate}
   	          
Among all listed entities, the reward for a certificate submitter is dynamic, while the remainder is distributed in fixed proportions.
   	          
The value ${Er}^{gl\_submitter}_j$ is set by the submitter itself. The mechanism is similar to the one adopted for provers. The certificate submitter is chosen among the forgers of the epoch. While issuing a block, a forger makes a bid for the reward he wants to receive for submitting a certificate. The forger that made the lowest bid has the first priority to submit a certificate at the end of the epoch. If it failed to submit during the prescribed period, the forger with the second lowest bid gets a chance to do this and so forth (see details in [\ref{sec:submission_mechanism} \nameref{sec:submission_mechanism}]). The ${Er}^{gl\_submitter}_j$ will be equal to the bid of the forger that eventually submitted a certificate to the mainchain, but no more than the maximum submitter reward, that is  ${sub\_max  \cdot  Er}^{gl}_j$, $sub\_max \in [0,1]$ is a system parameter.
   	           
After the deduction of ${Er}^{gl\_submitter}_j$ from the global pool, the residual rewards are distributed among all the other categories according to the following proportions:
\[{Er}^{gl}_j-{{Er}^{gl\_submitter}_j=Er}^{gl\_forgers}_j+{Er}^{gl\_refs}_j+{Er}^{gl\_scdev}_j+{Er}^{gl\_circuitdev}_j,\]
\[{Er}^{gl\_forgers}_j=fgs  \cdot ({Er}^{gl}_j-{Er}^{gl\_submitter}_j),\]
\[{Er}^{gl\_refs}_j=refs \cdot ({Er}^{gl}_j-{Er}^{gl\_submitter}_j),\]
\[{Er}^{gl\_scdev}_j=dev \cdot ({Er}^{gl}_j-{Er}^{gl\_submitter}_j),\]
\[{Er}^{gl\_circuitdev}_j=cdev \cdot ({Er}^{gl}_j-{Er}^{gl\_submitter}_j),\] 
where $\{fgs,refs,dev,cdev\} \in [0,1]$ are system parameters satisfying $fgs+refs +dev+cdev=1$. 
          	
Moreover, the forgers reward ${Er}^{gl\_forgers}_j$ is further divided equally among all forgers of the epoch:
\[{Er}^{gl\_forgers}_j=fgs  \cdot ({Er}^{gl}_j-{Er}^{gl\_submitter}_j)=\sum^F_{f=0}{}{Er}^{gl\_forgers(f)}_j,\]
\[{Er}^{gl\_forgers(f)}_j=\frac{{Er}^{gl\_forgers}_j}{F},\]
where $F$ is the total number of forgers that issued blocks in the epoch.  
          	  
The reward ${Er}^{gl\_refs}_j$ is divided among forgers that included mainchain block references into their blocks:
\[{Er}^{gl\_refs}_j=refs  \cdot ({Er}^{gl}_j-{Er}^{gl\_submitter}_j)=\sum^R_{r=0}{}{Er}^{gl\_forgers(r)}_j,\]
\[{Er}^{gl\_forgers(r)}_j=\frac{{Er}^{gl\_forgers}_j}{R},\]
where $R$ is the total number of mainchain block references included into all the blocks in the withdrawal epoch. ${Er}^{gl\_forgers(r)}_j$ is the amount of rewards paid for each MC reference. Note that one forger can get rewards for several MC references.  
\\~\\          	  
\textbf{Example 2}. Let us assume that ${Er}^{gl}_j=1000$ is an amount of fees collected into the global pool. Let also assume that $F=100$ forgers issued blocks in the epoch, $R=50$ is the number of MC block references, and ${Er}^{gl\_submitter}_j=100$ is the amount requested by the certificate submitter. Assume the following system parameters:  
          	  
\begin{itemize}
    \item  $sub\_max=0.5$;
    \item  $fgs=0.5$;
    \item  $refs=0.3$;
    \item  $dev=0.1$;
    \item  $cdev=0.1$.
\end{itemize}   
          	  
Then, the global rewards distribution is as follows:  \[{Er}^{gl}_j-{Er}^{gl\_submitter}_j=900,\]
\[{Er}^{gl\_forgers}_j=fgs \cdot 900=450,\]
\[{Er}^{gl\_forgers(f)}_j=\frac{{Er}^{gl\_forgers}_j}{F}=\frac{450}{100}=4.5,\]
\[{Er}^{gl\_refs}_j=refs  \cdot 900=270,\]
\[{Er}^{gl\_refs(f)}_j=\frac{{Er}^{gl\_refs}_j}{R}=\frac{270}{50}=5.4,\]
\[{Er}^{gl\_scdev}_j=dev \cdot 900=90,\]
\[{Er}^{gl\_circuitdev}_j=cdev \cdot  900=90.\]
\section{Conclusions and Future Directions}

We presented an incentive scheme for the Latus sidechain construction \cite{GKO20} that
facilitates sustainable maintenance and promotion of decentralization purely through transaction fees. In the paper we identified the main parties involved in sidechain operation and constructed a reward sharing scheme that maximizes overall efficiency of the system and keeps different entities in equilibrium.

We view our work as a first step in constructing a robust incentive scheme for a
sidechain system that would not possess its own native asset and, thus, cannot reward
maintainers with newly minted coins as it is done in most of the existing blockchain systems. There are many questions still open in this setting that are worth further research. One of the most interesting and challenging is to analyze the scheme using game-theoretic techniques and define optimal system parameters. It is important to find out what exact parameters should be used in a real world system to encourage maintainers to join and behave as expected. Finally, it is also interesting how to make parallelization of proofs generation more efficient and eliminate duplicate work by different provers.

\begin{appendices}
\section{Withdrawal Certificate Signing} \label{sec:AppendixA}  
 
\subsection{Motivation}

As it was briefly discussed in [\ref{sec:wcert} \nameref{sec:wcert}], a withdrawal certificate can be submitted to the mainchain only if at least half of the forgers of the withdrawal epoch signed it.

To understand the reason for having threshold signature, it is important to remember the protocol followed by the mainchain in the case of receiving multiple certificates from the same sidechain in the same epoch. In such a situation, the mainchain will select a certificate with the highest quality parameter. In the Latus sidechain, we use the number of sidechain blocks that are part of the withdrawal epoch as the quality. For example, in the case of a fork in the sidechain resulting in two different branches at the end of the epoch, the quality of the two certificates will be the number of blocks in the corresponding branches. In such a scenario, the mainchain will implicitly accept the certificate associated with the longest branch. The reason for having this kind of the selection process is quite obvious, but what is not immediately obvious is that such a scheme may open an opportunity for a malicious actor to submit a certificate committing a valid but unknown state in the mainchain (unknown to the rest of the sidechain network). That may be possible, for example, for a slot leader having the right to issue the last block in the withdrawal epoch by
extending the longest chain, keeping the block hidden, proving the hidden block transitions and finally submitting the certificate into the mainchain. No one else in the sidechain will be able to create a certificate with the higher quality, and the mainchain will then consider as final the certificate committing a state unknown to anyone except the malicious actor.

To prevent such situation, we require at least half of the forgers who issued
blocks in the epoch, to confirm it by signing. With such a procedure, a malicious actor will not be able to perform the attack unless he controls the majority of the forgers of the withdrawal epoch.

\subsection{Signing protocol}

\begin{figure}[htbp]
	\centering
	\includegraphics[trim={1cm 9cm 4cm 5.42cm}, clip,width=1\columnwidth] {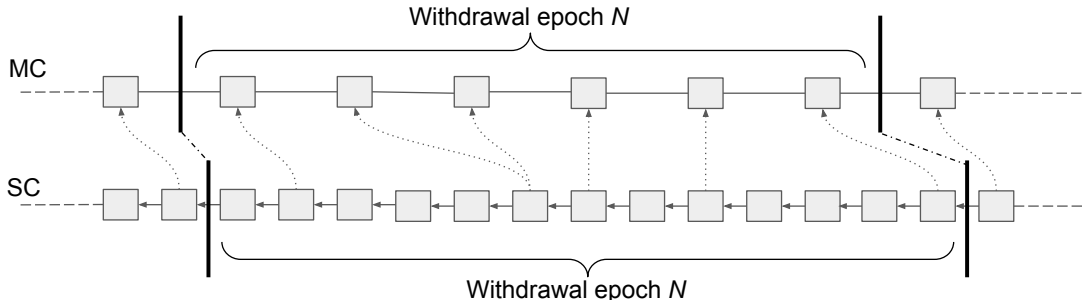}
	\caption{A withdrawal epoch in the sidechain is defined by a range of MC blocks}
	\label{fig:A_1}
\end{figure}

According to the consensus protocol of the Latus sidechain, the length of a withdrawal epoch, as well as a certificate submission period, are defined by mainchain block references. The withdrawal epoch \textit{N }in the sidechain includes all slots between the slot with the block that contains reference to the MC block that begins the epoch in the mainchain and the slot containing reference to the MC block that begins the following epoch (see Fig.\ref{fig:A_1}).  

A withdrawal certificate for the epoch \textit{N }can be constructed deterministically after the last block of epoch \textit{N} is issued. Every forger does this independently and signs the certificate. If forgers have the same view of the sidechain, they will sign the same certificate and implicitly agree on the same state. The signature is propagated to other forgers immediately. Importantly, the signatures are not submitted to the blockchain, they are collected off-chain.   
 
When submitting the certificate, the submitter collects all signatures and creates one aggregated signature that is attached to the certificate. The certificate's proof will be valid only if at least half of the forgers signed the certificate and only if the quality parameter is equal to the real number of forgers that were part of the epoch. 
\end{appendices}
\begin{appendices}
\section{Cross-Chain Transaction Fees} \label{sec:AppendixB}  
 
The rewards for referencing mainchain blocks consist mostly of the fees paid by cross-chain transactions included into referenced mainchain blocks. There are two types of cross-chain transactions that may be originated in the mainchain: forward transfer (FT) and backward transfer request (BTR). The problem is that once such a transaction is included into the mainchain block, there is no way for a sidechain forger to eliminate its processing, even if the fee is zero, because he is obliged to process all cross-chain transfers present in the referenced block. 
 
To resolve this issue, we require paying of at least the minimal fee for creating of a cross-chain transaction in the mainchain that is calculated as the median value of transaction fees in the sidechain during the previous epoch multiplied on the complexity coefficient (FTs and BTRs might be more complicated than regular transactions). This rule is enforced by the mainchain. To make the mainchain aware about the minimal fee required by the sidechain, a withdrawal certificate will explicitly specify it. In turn, the value of the fee is enforced by the certificate proof, so the submitter cannot manipulate it. It is important to mention that the mainchain does not know how this number is calculated, it just disallows cross-chain transactions that pay less than the minimal fee. 
 
Technically, the fees should be paid as a separate output transferred to the sidechain. It is especially important for the BTRs, because it might happen that a BTR tries to withdraw nonexistent UTXO in the sidechain, so there would be no way to tax the requester unless he directly sends a fee to the sidechain. So the BTR should include a forward transfer to pay the fee. The amount of this transfer will be bounded by the minimal fee.  
\end{appendices}
  
\begin{appendices}
\section{Restrictions for cross-chain transfers} \label{sec:AppendixC}  
  
\subsection{Limit on backward transfers per withdrawal certificate} 
 
Given that mainchain blocks are limited by size, that imposes a limit on withdrawal certificate size and, consequently, on the number of backward transfers included into the certificate. Let us assume that the maximal number of BTs that can be processed is $N$. Note that this limit affects both withdrawal requests on the sidechain side and backward transfer requests on the mainchain side, each of these operations turns into a BT in the certificate.

But if we simply implement the first-come first-served (FCFS) strategy, where the first \textit{N }BTs are included into the sidechain, while others are rejected, it might lead to censorship attacks on withdrawals. E.g., a malicious forger having several slots at the beginning of the epoch might fill up all the space for BTs. Situation gets even worse if the sidechain is completely corrupted (having more than 50\% of malicious forgers), in this case an adversary might continuously prevent withdrawals by placing his own BTs first. Even BTRs on the mainchain side that were initially introduced to prevent censorship of withdrawals would not help because they are also subjected to BTs limit, and if the adversary creates his BTs first, there would be no more space for BTRs from the mainchain. 
 
To solve this problem and provide a way to create BTRs even in the case of a corrupted sidechain, we introduce a mechanism that unlocks the ability to include withdrawal requests gradually throughout the epoch. Given that there are \textit{N} BTs allowed per withdrawal epoch and that the sidechain epoch is bounded by \textit{M} mainchain block, we suggest the following mechanism:  
  
\begin{enumerate} 
    \item  $N \cdot f$ spots for BTs are distributed according to the FCFS strategy, where $f \in [0,1]$ is a system parameter defining the portion of spots for FCFS;
    \item  $N \cdot (1-f)$ spots for BTs are unlocked gradually as MC block references are included into the sidechain block. Each MC reference unlocks $\frac{N \cdot (1-f)}{M}$ new spots for BTs.
\end{enumerate} 
  
Note that in this scheme backward transfer requests originated in the mainchain have preference upon withdrawal requests created in the sidechain in the case they are competing for the same spots in a SC block. That is, if all FCFS spots are already consumed and a new SC block includes a MC block reference that unlocks the following portion of BTs spots, the BTRs included into this MC block reference are processed first and only then, if there are spare spots left, the SC withdrawal requests can be included into this SC block. In this case, there is always a possibility to create a backward transfer request that will be processed, even if the sidechain is totally corrupted.
   
This scheme implies that the number of BTRs per mainchain block is bounded by $\frac{N  \cdot  (1-f)}{M}$ value, because we cannot guarantee that more than this number will be eventually processed by the sidechain. Note that a forward transfer may also turn into a backward transfer in the same epoch, in the case FT failed to be processed in the sidechain. It means the number of both BTRs and FTs should not be more than $\frac{N \cdot (1-f)}{M}$ .
  
\subsection{Limit on cross-chain transactions per mainchain block}
  
Given that the mainchain itself does not know how many FTs and BTRs can be processed by the sidechain, it should be explicitly specified upon sidechain registration. We suggest two additional values: \textit{maxForwardTransfersPerBlock} and \textit{maxBackwardTransferRequestsPerBlock} that are provided upon registration. Then, the mainchain prohibits including into a MC block more than a specified number of cross-chain transfers for a particular sidechain. This is needed to prevent SC not to  have the computational power to process all cross-chain transactions.  
  
Note that these values can actually be smaller than $\frac{N  \cdot (1-f)}{M}$, but their sum cannot be larger.  
\end{appendices}
 \begin{appendices}

\section{State-transition proofs tree} \label{sec:AppendixD}  

As it was briefly described in [\ref{sec:distr_proof_gen} \nameref{sec:distr_proof_gen}], a state transition proof for a block is constructed recursively by merging proofs of basic transitions. A tree of proofs is a tree whose leaves represent a set of sidechain transactions. Given that there are many ways to construct the tree from a list of leaves, it is very important for provers and forgers to establish a canonical way to do this, because otherwise they may end up building inconsistent proofs that cannot be merged. In this section, we discuss in more detail how exactly the tree is constructed.

\subsection{Merkle Mountain Range trees}

We adopt the Merkle Mountain Range construction (MMR)  \cite{TMMR18,MMRDOC} to build an efficient tree of proofs. The MMR is a Merkle tree that allows an efficient appending operation. Note that the tree of proofs is somewhat similar to a Merkle tree, with the difference that instead of a hash function we use a SNARK prover.   

The basic idea of MMR is that, given a list of leaves, it produces a set of largest possible perfect\footnote{Perfect binary tree is a tree with $2^i$ leaves for some ($0 \leq i$). For example: 1, 2, 4, 8, 16, etc.} binary trees, called mountains. Each mountain has a peak; those picks are then combined into a single MMR tree.   

\begin{figure}[htbp]
	\centering
	\includegraphics[trim={2.5cm 10.5cm 4cm 3.5cm}, clip,width=1\columnwidth] {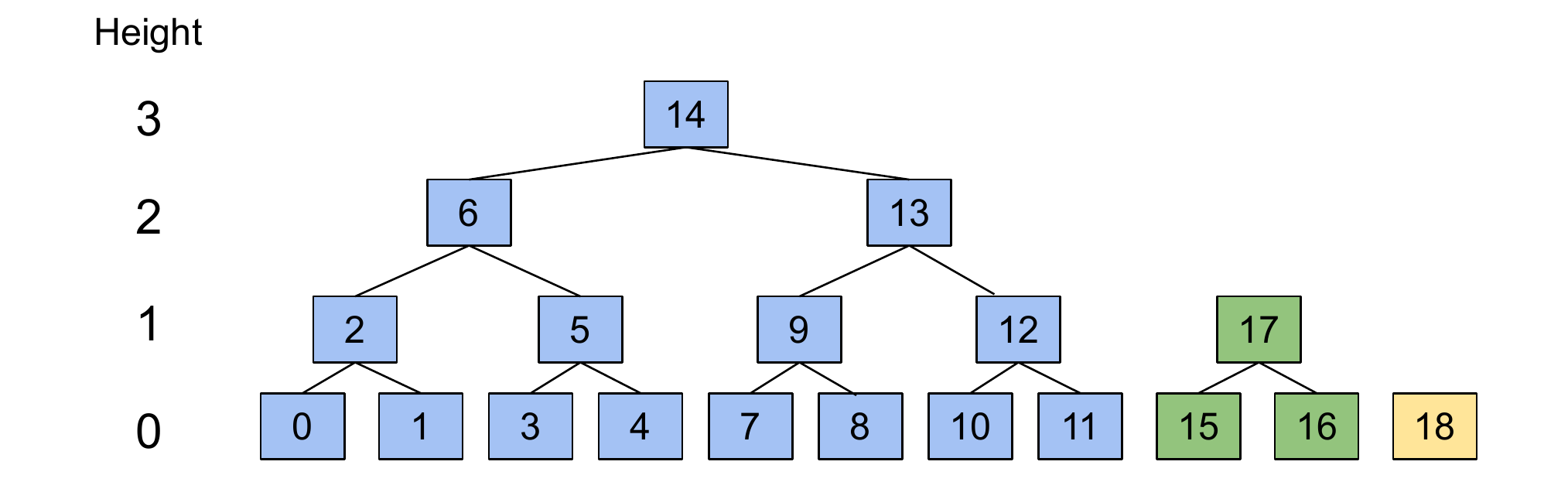}
	\caption{Mountains in the Merkle Mounting Range tree. Different mountains are highlighted with different colors.}
	\label{fig:D_1}
\end{figure}

Let us consider an example on Fig.~\ref{fig:D_1}. There are eleven leaves that allows us to construct 3 mountains with peaks at nodes 14, 17 and 18. Each mountain is a perfect binary tree where the number of leaves is a power of 2. 

The mountains are constructed by joining the sibling trees of the same size. It is done by traversing leaves from left to right and adding a parent node as soon as 2 corresponding children exist (nodes on Fig.~\ref{fig:D_1} are numbered in order of traversal). Note that the mountains are decreasing in size: each following subtree has at most half of the leaves.  
 
Traversal can also be represented as a flat list:

\setlength\parindent{34pt}
\textbf{Position   0 \ 1 \ 2 \ 3 \ 4 \ 5 \ 6 \ 7 \ 8 \ 9 \ 10 \ 11 \ 12 \ 13 \ 14 \ 15 \ 16 \ 17 \ 18} 

\begin{figure}[htbp]
	\centering
	\includegraphics[trim={4.5cm 10.5cm 4cm 1.5cm}, clip,width=1\columnwidth] {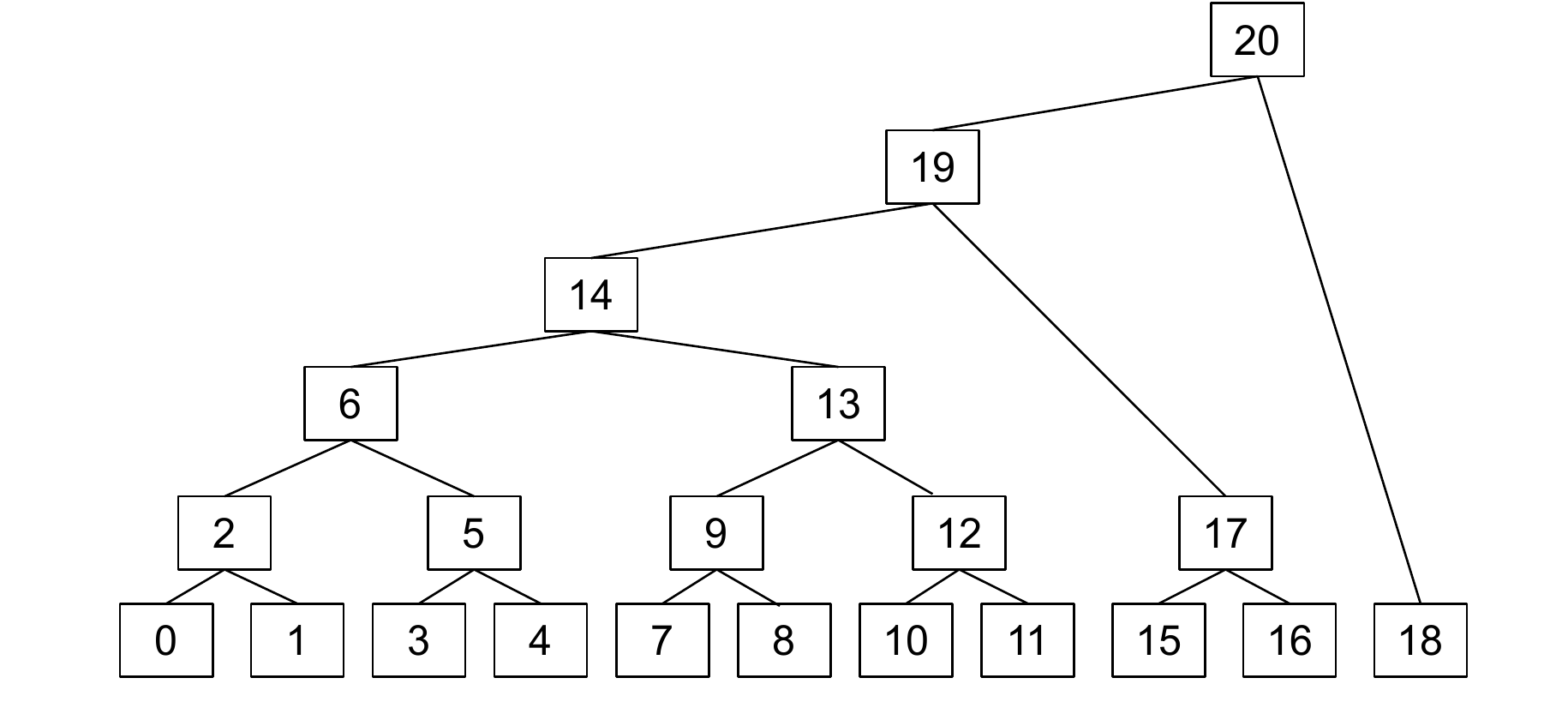}
	\caption{Full Merkle Mounting Range tree}
	\label{fig:D_2}
\end{figure}

\setlength\parindent{34pt}
\textbf{Height \ \ 0 \ 0 \ 1 \ 0 \ 0 \ 1 \ 2 \ 0 \ 0 \ 1 \ \ 0 \ \ 0 \ \ \ 1 \ \ 2 \ \ \ 3 \ \ 0 \ \ \ 0 \ \ 1 \ \ \ 0}   
 \\~\\
 Once the mountains are constructed, they are combined into a single tree as follows: 
 
\begin{enumerate}
    \item A new node is added combining two leftmost peaks.
    \item Step 1 is repeated until a single peak left (see Fig.~\ref{fig:D_2}).
\end{enumerate}
 
In general, MMR trees allow efficient incremental updates such as appending, removing last leaf, pruning leaves.     
 
\subsection{Using Merkle Mountain Range trees for recursive proof generation}   
  
We adopt MMR trees to construct a tree of proofs. Assuming leaves of the tree are basic proofs for transactions broadcasted in a transactions proposal, each prover and forger follows the rules for MMR tree construction. The forger includes into the block, as a state transition proof, the peak of the MMR tree or, if the tree was not completely built, it includes the leftmost peak.
  
\begin{figure}[htbp]
	\centering
	\includegraphics[trim={3cm 9.5cm 2.8cm 1.5cm}, clip,width=1\columnwidth] {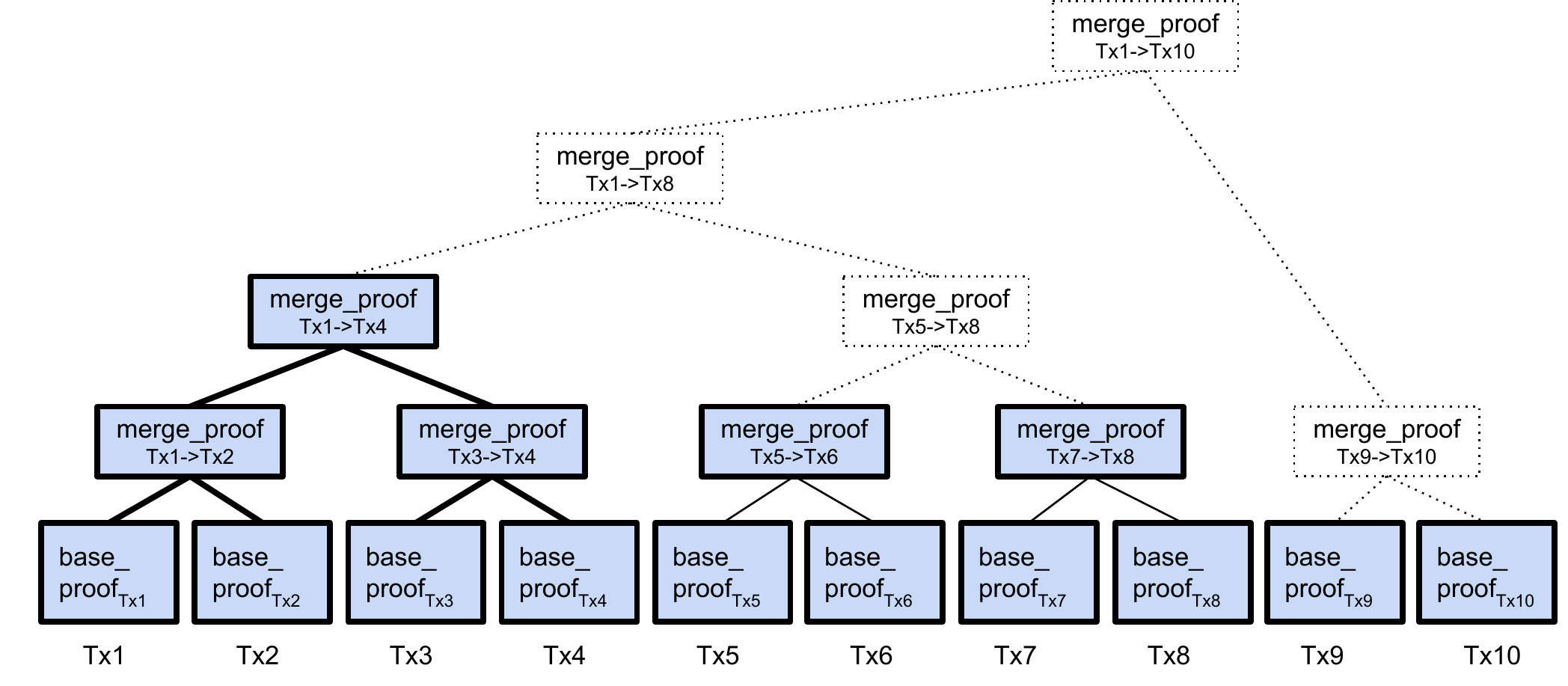}
	\caption{MMR tree of proofs}
	\label{fig:D_3}
\end{figure}

For instance, let us assume the following transactions proposal: $\{Tx1,\ ...,\ Tx10\}$. For this proposal provers have to construct the MMR tree as on Fig.~\ref{fig:D_3}. Let us also assume that until the end of the slot forgers were are to generate only a subset of the required tree (highlighted with color on Fig.~\ref{fig:D_3}).
  
According to the rules, the forger has to include $merge\_proof_{Tx1 \rightarrow   Tx4}$ as a state transition proof into the block, because it is the leftmost peak. Note that other available peaks are $merge\_proof_{Tx5 \rightarrow   Tx6}$, $merge\_proof_{Tx7 \rightarrow  Tx8}$, $base\_proof_{Tx9}$, and $base \_proof_{Tx10}$, but if the forger uses one of them, the preceding tree of proofs would be wasted. The default behavior of the forger is to include the first available leftmost peak, even if it is not the largest subtree. 
   
Such a construction provides one important property: it allows efficient transition of already generated proofs to the following slot. For instance, if the forger uses $merge\_proof_{Tx1 \rightarrow  Tx4}$ from Fig.~\ref{fig:D_3} as a block transition proof, the forger of the following slot can broadcast a transactions proposal $\{Tx5,Tx6,\ ...,\ Tx15\}$, in which case already generated, but not used, proofs for the transactions $\{Tx5,Tx6,\ ...,\ T10\}$ may be reused.   
\end{appendices}

\bibliographystyle{plain}
\bibliography{references}
\end{document}